\documentclass[12pt,english]{article}

\usepackage{natbib}
\usepackage{url}
\makeatletter
\def\url@leostyle{%
  \def\UrlFont{\sf}}{\def\UrlFont{\small\ttfamily}}
\makeatother
\usepackage{geometry}
\usepackage{fancyhdr}
\usepackage[font=footnotesize,labelfont=bf]{caption}

\usepackage{authblk}

\usepackage{hyperref}

\input{Qcircuit}

\numberwithin{equation}{section}

\renewcommand{\Qcircuit}[1][0em]{\xymatrix @*=<#1>}

\usepackage{graphicx}
\usepackage{babel}
\usepackage{amsmath}
\usepackage{amsfonts}
\usepackage{mathrsfs}

\begin{document}

\title{On the Necessity of Entanglement for the Explanation of
  Quantum Speedup\thanks{This paper has been adapted
    from Chapter 3 of \citet[]{cuffaro2013}. I am indebted to Wayne
    Myrvold for discussion and for his comments and criticisms on
    earlier drafts.}}
\author{Michael E. Cuffaro}
\affil{The University of Western Ontario, Department of Philosophy}

\maketitle

\thispagestyle{empty}
\pagenumbering{roman}
\tableofcontents
\newpage
\pagenumbering{arabic}

\section{Introduction}
\label{s:intro}

The significance of the phenomenon of quantum entanglement\textemdash
wherein the most precise characterisation of a quantum system composed
of previously interacting subsystems does not necessarily include a
precise characterisation of those subsystems\textemdash has been at
the forefront of the debate over the conceptual foundations of quantum
theory, almost since that theory's inception. It is \emph{the}
distinguishing feature of quantum theory, for some
\citep[]{schrodinger1935}.\footnote{For some more recent speculation
  on the the distinguishing feature(s) of quantum mechanics, see,
  for instance, \citet[]{clifton2003,myrvold2010}.} For others, it is
evidence for the incompleteness of that theory
\citep*[]{epr1935}.\footnote{For further discussion, and for
  Einstein's later refinements of the Einstein-Podolsky-Rosen (EPR)
  paper's main argument, see \citet[]{howard1985}.} For yet others,
the possibility of entangled quantum systems implies that physical
reality is essentially non-local \citep[]{stapp1997}.\footnote{For
  responses to Stapp's view and for further discussion, see:
  \citet[]{unruh1999,mermin1998,stapp1999}.} For almost all, it has
been, and continues to be, an enigma requiring a solution.

For most of the history of quantum theory, serious investigation into
the significance and implications of entanglement has been conducted
mainly by philosophers of physics and by a few philosophically-minded
theoretical and experimental physicists interested in foundational
issues. With the advent of quantum information theory, this has begun
to change. In quantum information theory, quantum mechanical systems
are utilised to implement communications protocols and computational
algorithms that are faster and more efficient than any of their known
classical counterparts. Because it is almost surely the case that one
or more of the fundamental distinguishing aspects of quantum mechanics
is responsible for this `quantum advantage', quantum information
theory has precipitated an explosion of physical research into the
traditionally foundational issues of quantum theory.

Of the many and varied applications of quantum information theory,
perhaps the most fascinating is the sub-field of quantum
computation. In this sub-field, computational algorithms are designed
which utilise the resources available in quantum systems in order to
compute solutions to computational problems with, in some cases,
exponentially fewer resources than any known classical algorithm. A
striking example of this so-called `quantum speedup' is Shor's
algorithm \citep{shor1997} for factoring integers. A basic
distinction, in computational complexity theory, is between those
computational problems that are amenable to an \emph{efficient}
solution in terms of time and space resources, and those that are
not. Easy (or `tractable', `feasible', `efficiently solvable', etc.)
problems are those which involve resources bounded by a polynomial in
the input size, $n$ ($n^c$ time steps, for instance, where $c$ is some
constant). Hard problems are those which are not easy; they are those
problems whose solution requires resources that are `exponential' in
$n$, i.e., that grow faster than any polynomial in $n$.\footnote{As
  this class of problems includes those solvable in, for instance,
  $n^{\log n}$ steps, this convention abuses, somewhat, the term
  exponential, hence my use of inverted commas.}$^,$\footnote{As we
  will discuss in more detail later, the easy-hard distinction is
  not meant to reflect any deep mathematical truth about the nature
  of computational algorithms, but is rather meant as a practical
  characterisation of what we normally associate with efficiency.}
The factoring problem is believed to be hard, classically, and indeed,
much of current Internet security relies on this fact. Shor's quantum
algorithm for factoring integers, however, makes the factoring problem
efficiently solvable.

While the fact of quantum computational speedup is almost beyond
doubt,\footnote{Just as with other important problems in computational
  complexity theory, such as the \textbf{P} = \textbf{NP} problem,
  there is currently no proof, though it is very strongly suspected to
  be true, that the class of problems efficiently solvable by a
  quantum computer is larger than the class of problems efficiently
  solvable by a classical computer.} the source of quantum speedup is
still a matter of debate. Candidate explanations of quantum speedup
range from the purported ability of quantum computers to perform
multiple function evaluations simultaneously
\citep[][]{deutsch1997,duwell2004,hewittHorsman2009},\footnote{For
  criticisms of the version of this view that takes this parallel
  computation to occur in many parallel universes, see, for instance,
  \citet[]{steane2003,duwell2007,cuffaro2012}.} to the purported
ability of a quantum computer to compute a global property of a
function without evaluating \emph{any} of its values
\citep[e.g.][]{steane2003,bub2010}.

In most of these candidate explanations, the fact that quantum
mechanical systems can sometimes exhibit \emph{entanglement} plays an
important role. On A.M. Steane's view, for instance, quantum
entanglement allows one to manipulate the correlations between the
values of a function without manipulating those values themselves. For
proponents of the many worlds explanation, on the other hand, though
they consider computational worlds to be the main component in the
explanation of quantum speedup, they nevertheless view entanglement as
indispensable to its analysis \citep[889]{hewittHorsman2009}. It is
thus somewhat disconcerting that recent physical research seems to
suggest that entanglement, rather than being indispensable, may be
irrelevant to the general explanation of quantum speedup.

Logically, entanglement may play the role of either a necessary or a
sufficient condition (or both) in an overall explanation of quantum
speedup. I address the question of whether entanglement may be said to
be a sufficient condition elsewhere \citep[]{cuffaro2013}. As for
the assertion that entanglement is a necessary component in the
explanation of speedup, this seems, prima facie, to be
supported\footnote{What I take to be supported by
  \citeauthor[]{jozsa2003}'s result is the claim that entanglement
  is required in order \emph{to explain} quantum speedup. As we will
  discuss further in \textsection \ref{s:net}, this is
  distinct from the claim that one requires an entangled quantum
  state in order to achieve quantum speedup.} by a result due to
\citet*[]{jozsa2003}, who prove that for quantum algorithms which
utilise \emph{pure} states, ``the presence of multi-partite
entanglement, with a number of parties that increases unboundedly with
input size, is necessary if the quantum algorithm is to offer an
exponential speed-up over classical computation''
\citeyearpar[p. 2014]{jozsa2003}. When we consider quantum algorithms
which utilise \emph{mixed} states, however, then there appear to be
counterexamples to the assertion that one must appeal to quantum
entanglement in order to explain quantum speedup. In particular,
\citet[]{biham2004} have shown that it is possible to achieve a modest
(sub-exponential) speedup using unentangled mixed states. Further,
\citet[]{datta2005,datta2008} have shown that it is possible to
achieve an exponential speedup using mixed states that contain only a
vanishingly small amount of entanglement. In the latter case, further
investigation has suggested to some that quantum correlations
\emph{other than} entanglement may be playing a more important
role. One quantity in particular, \emph{quantum discord}, appears to
be intimately connected to the speedup that is present in the
algorithm in question. In light of these results, it is tempting to
conclude that it is not necessary to appeal to entanglement at all in
order to explain quantum computational speedup and that the
investigative focus should shift to the physical characteristics of
quantum discord or some other such quantum correlation measure
instead.

In this paper I will argue that this conclusion is premature and
misguided, for as I will show below, there is an important sense in
which entanglement can indeed be said to be necessary for the
explanation of the quantum speedup obtainable from both of these
mixed-state quantum algorithms.

I will proceed as follows. After introducing the concept of
entanglement and how it is quantified in \textsection
\ref{s:prelim}, I introduce the \emph{necessity of entanglement
  for explanation thesis} in \textsection \ref{s:net}. In
\textsection \ref{s:deq}, I show how what looks like a
counter-example to the necessity of entanglement for explanation
thesis for pure states\textemdash the fact that certain important
quantum algorithms can be expressed so that their states are never
entangled\textemdash is instead evidence for this thesis. Then, in
\textsection\ref{s:mix}, I examine the more serious challenges to
the necessity of entanglement for explanation thesis posed by the
cases of sub-exponential speedup with unentangled mixed states
(\textsection\ref{s:mixdj}) and exponential speedup with mixed
states containing only a vanishingly small quantity of entanglement
(\textsection\ref{s:oneq}).

Starting with the first type of counter-example, I begin by arguing
that pure quantum states should be taken to provide a more fundamental
representation of quantum systems than mixed states. I then show that
when one considers the initially mixed state of the quantum computer
as representing the space of its possible pure state preparations, the
speedup obtainable from the computer can be seen as stemming from the
fact that the quantum computer evolves some of these possible pure
state preparations to entangled states\textemdash that the quantum
speedup of the computer can be seen as arising from the fact that it
implements an entangling transformation.

As for the second type of counter-example, where exponential speedup
is achieved with only a vanishingly small amount of entanglement, and
where it is held by some that another type of non-classical
correlation, quantum discord, is responsible for the speedup of the
quantum computer: I argue that, first, it is misleading to
characterise discord as indicative of non-classical correlations. I
then appeal to recent work done by \citet[]{fanchini2011},
\citet[]{brodutch2011}, and \citet[]{cavalcanti2011} who show,
respectively, that when one considers the `purified' state
representation of the quantum computer, there is a conservation
relation between discord and entanglement, and indeed that there is
just as much entanglement in such a representation as there is discord
in the mixed state representation; that entanglement must be shared
between two parties in order to bilocally implement any bipartite
quantum gate; and that entanglement is directly involved in the
operational definition of quantum discord.

Given \citeauthor[]{jozsa2003}'s proof of the necessary presence of
an entangled state for exponential speedup using pure states, and
given the fundamentality of pure states as representations of quantum
systems, the burden of proof is upon those who would deny the
necessity of entanglement for explanation thesis to show either by
means of a counter-example or by some other more principled method
that it is false. Neither of the counter-examples discussed in this
paper succeeds in doing so. We should conclude, therefore, that the
necessity of entanglement for explanation thesis is true.

\section{Preliminaries}
\label{s:prelim}

\subsection{Quantum entanglement}
\label{s:ent}

Consider the following representation of the joint state of two
qubits:\footnote{A qubit is the basic unit of quantum information,
  analogous to a classical bit. It can be physically realised by any
  two-level quantum mechanical system. Like a bit, it can be ``on'':
  $| 1 \rangle$ or ``off'': $| 0 \rangle$, but unlike a bit it can
  also be in a superposition of these values.}
$$| \psi \rangle = | 0 \rangle \otimes | 0 \rangle + | 0 \rangle
\otimes | 1 \rangle + | 1 \rangle \otimes | 0 \rangle + | 1 \rangle
\otimes | 1 \rangle.$$
This expression for the overall state of the system represents the
fact that the two qubits are in an equally weighted superposition of
the four joint states (a)-(d) below:
\vspace{1em}
\begin{center}
\begin{tabular}{lrr}
& $q_1$ & $q_2$ \\
(a) & $| 0 \rangle$ & $| 0 \rangle$ \\
(b) & $| 0 \rangle$ & $| 1 \rangle$ \\
(c) & $| 1 \rangle$ & $| 0 \rangle$ \\
(d) & $| 1 \rangle$ & $| 1 \rangle$. \\
\end{tabular}
\end{center}
\vspace{1em}
This particular state is a \emph{separable} state, for it can,
alternatively, be expressed as a product of the pure states of its
component systems, as follows:
$$| \psi \rangle = (| 0 \rangle + | 1 \rangle) \otimes (| 0 \rangle +
| 1 \rangle).$$

Not all quantum mechanical states can be expressed as product states
of their component systems, and thus not all quantum mechanical states
are separable. Here are four such `entangled' states:\footnote{From
  now on, I will usually, for brevity, omit the tensor product symbol
  from expressions for states of multi-particle systems; i.e., $|
  \alpha\beta \rangle$ and $| \alpha \rangle| \beta \rangle$ should
  be understood as shorthand forms of $| \alpha\rangle\otimes| \beta
  \rangle$.}
\begin{align*}
|\Phi^+\rangle = \frac{| 00 \rangle + | 11 \rangle}{\sqrt{2}}
\nonumber \\
|\Phi^-\rangle = \frac{| 00 \rangle - | 11 \rangle}{\sqrt{2}}
\nonumber \\
|\Psi^+\rangle = \frac{| 01 \rangle + | 10 \rangle}{\sqrt{2}}
\nonumber \\
|\Psi^-\rangle = \frac{| 01 \rangle - | 10 \rangle}{\sqrt{2}}.
\end{align*}
The skeptical reader is encouraged to convince himself that it is
impossible to re-express any of these states as a product state of two
qubits. They are called the Bell states, and I will refer to a pair of
qubits jointly in a Bell state as a Bell pair.\footnote{These are
  also sometimes referred to as `EPR pairs'. EPR stands for
  Einstein, Podolsky, and Rosen. In their seminal
  \citeyear[]{epr1935} paper, EPR famously used states analogous to
  the Bell states to argue that quantum mechanics is incomplete.}
Maximally entangled states,\footnote{Note that not all entangled
  states are maximally entangled states. We will discuss this in more detail
  shortly.} such as these, completely specify
the correlations between outcomes of experiments on their component
qubits without specifying anything regarding the outcome of a single
experiment on one of the qubits. For instance, in the singlet state
(i.e., $|\Psi^-\rangle$), outcomes of experiments on the first and
second qubits are perfectly anti-correlated with one another. If one
performs, say, a $\hat{z}$ experiment on one qubit of such a system,
then if the result is $|0\rangle$, a $\hat{z}$ experiment on the other
qubit will, with certainty, yield an outcome of $|1\rangle$, and vice
versa. In general, the expectation value for joint measurements on the
two qubits is given by $- \hat{m} \cdot \hat{n} = - \cos\theta,$ where
$\hat{m}, \hat{n}$ are unit vectors representing the orientations of
the two experimental devices, and $\theta$ is the difference in these
orientations. Any single $\hat{z}$ experiment on just one of the two
qubits, however, will yield $|0\rangle$ or $|1\rangle$ with equal
probability.

We will put to one side the question of the \emph{physical
  significance} of quantum entanglement. I discuss this at greater
length in Chapters 4 and 5 of \citet[]{cuffaro2013}. For the purposes
of this paper it is most appropriate to give as minimal and
uncontroversial a characterisation of entanglement as possible.

\subsection{Entangled mixed states}
\label{s:mixent}

The concepts of separability and of entanglement are also applicable
to so-called `mixed states'. To illustrate the concept of a mixed
state, imagine that one draws a ball from an urn into which balls of
different types have been placed, and that the probability of drawing
a ball of type $i$ is $p_i$. Corresponding to the outcome $i$, we then
prepare a given system $S$ in the pure state $| \psi_i \rangle$,
representable by the density operator $\rho^S_i = | \psi_i
\rangle\langle \psi_i |$. After preparing $\rho^S_i$, we then discard
our record of the result of the draw. The resulting state of the
overall system will be the mixed state:
\begin{equation}
\label{eqn:mix}
\rho = \sum_i p_i \rho_i^S.
\end{equation}

A \emph{mixed state} is \emph{separable} if it can be expressed as a
mixture of pure product states, and \emph{entangled} otherwise. In
general, determining whether a mixed state of the form
\eqref{eqn:mix} is an entangled state is difficult, because in
general the decomposition of mixtures is non-unique. For instance, the
reader can verify that a mixed state represented by the density
operator $\rho$, prepared as a mixture of pure states in the following
way:
$$\rho = \frac{3}{4}| 0 \rangle\langle 0 | + \frac{1}{4}| 1
\rangle\langle 1 |,$$
can also be equivalently prepared as:
$$\rho = \frac{1}{2}| \psi \rangle\langle \psi | + \frac{1}{2}| \phi
\rangle\langle \phi |,$$
where
$$| \psi \rangle \equiv \sqrt{\frac{3}{4}}| 0 \rangle +
\sqrt{\frac{1}{4}}| 1 \rangle, \quad | \phi \rangle \equiv
\sqrt\frac{3}{4}| 0 \rangle - \sqrt\frac{1}{4}| 1 \rangle.$$
This is so because both state preparations yield an identical density
matrix representation (in the computational basis); i.e.,:
\begin{align*}
\left(
\begin{matrix}
  3/4 & 0 \\
  0 & 1/4
\end{matrix}
\right).
\end{align*}
As we will see in more detail later, a system that is prepared as a
mixture of entangled states will sometimes yield the same density
operator representation as a system prepared as a mixture of pure
product states.

\subsection{Quantifying entanglement}

Entanglement is a potentially useful resource for quantum information
processing. \citet[]{masanes2006} has shown, for instance, that for
any non-separable state $\rho$, some other state $\sigma$ is capable
of having its teleportation fidelity enhanced by $\rho$'s
presence.\footnote{The teleportation fidelity \citep[cf.][\textsection
  9.2.2]{nielsenChuang2000} is a measure of the `closeness' of the
  input and output states in the teleportation protocol
  \citep[cf.][]{bennett1993}.} Given this, it is useful to be able to
quantify the amount of entanglement contained in a given state. In
order to do this, we employ so-called entanglement measures. Using
such measures, it is easy to see, for instance, that the state
\begin{equation}
\label{eqn:notmaxent}
| \phi \rangle = \sqrt\frac{1}{3}| 01 \rangle + \sqrt\frac{2}{3}| 10
\rangle,
\end{equation}
though entangled, is not a maximally entangled state (unlike the Bell
states we encountered in \textsection \ref{s:ent}, which are maximally
entangled). This is explained in more detail in \citet[]{cuffaro2013},
where a description of various entanglement measures is also
given. This can also be found in \citet[]{plenio2007}.

\subsection{Purification}
\label{s:purify}

Every mixed state can be thought of as the result of taking the
partial trace of a pure state acting on a larger Hilbert space. In
particular, for a mixed state $\rho_A$ acting on a Hilbert space
$\mathcal{H}_A$, with spectral decomposition $\sum_k p_k | k
\rangle\langle k |$ for some orthonormal basis $\{| k \rangle\}$, a
purification (in general non-unique) of $\rho_A$ may be given by $$|
\psi_{AB} \rangle = \sum_k \sqrt{p_k} | k_A \rangle \otimes | k_B \rangle
\in \mathcal{H}_A \otimes \mathcal{H}_B,$$ where $\mathcal{H}_B$ is a
copy of $\mathcal{H}_A$. We then have $\rho_A = \mbox{tr}_B(|
\psi_{AB} \rangle\langle \psi_{AB} |)$, with  $| \psi_{AB} \rangle$ an
entangled state.

\section{Entanglement in the quantum computer}
\label{s:qc}

\subsection{The Deutsch-Jozsa algorithm}
\label{s:djalgo}

Deutsch's problem \citep[]{deutsch1985} is the problem to determine
whether a given function $f:\{0,1\}\to\{0,1\}$ is constant or
balanced. Such a function is constant if it produces the same output
value for each of its inputs; it is balanced if the output of one half
of the inputs is the opposite of the output of the other half. Thus,
the constant functions from $\{0,1\}\to\{0,1\}$ are $f(x) = 0$ and
$f(x) = 1$; the balanced functions are the identity and bit-flip
functions.

A generalised version of this problem enlarges the class of functions
under consideration so as to include all of the functions
$f:\{0,1\}^n\to\{0,1\}$. Its quantum solution is given by the
Deutsch-Jozsa algorithm \citep[]{deutsch1992}. In
\citeauthor[]{cleve1998}'s improved version \citep[]{cleve1998}, the
algorithm begins by initialising the quantum registers of the computer
to $| 0^n \rangle| 1 \rangle$, after which we apply a Hadamard
transformation\footnote{The Hadamard transformation (also called a
  Hadamard `gate') takes $| 0 \rangle$ to $\frac{| 0 \rangle + | 1
  \rangle}{\sqrt 2}$ and $| 1 \rangle$ to $\frac{| 0 \rangle - | 1
  \rangle}{\sqrt 2}$ and vice-versa.} to all $n + 1$ qubits, so
that:
\begin{align}
| 0^n \rangle| 1 \rangle & \xrightarrow{H} \left(\frac{1}{2^{n/2}}(| 0
\rangle + | 1 \rangle)^n \right )\left(\frac{| 0 \rangle - | 1
  \rangle}{\sqrt 2}\right) \nonumber \\
\label{eqn:puredj_pre}
& = \left (\frac{1}{2^{n/2}}\sum_x^{2^n-1}| x \rangle
\right)\left(\frac{| 0 \rangle - | 1 \rangle}{\sqrt 2}\right).
\end{align}
The unitary transformation,
\begin{equation}
\label{eqn:puredj_uni}
U_f(| x \rangle | y \rangle) =_{df} | x \rangle | y \oplus f(x)
\rangle ,
\end{equation}
is then applied, which has the effect:\footnote{Given the state $| x
  \rangle(| 0 \rangle - | 1 \rangle)$ (omitting normalisation
  factors for simplicity), note that when $f(x)=0$, applying $U_f$
  yields $| x \rangle(| 0 \oplus 0 \rangle - | 1 \oplus 0 \rangle) =
  | x \rangle(| 0 \rangle - | 1 \rangle)$; and when $f(x) = 1$,
  applying $U_f$ yields $| x \rangle(| 0 \oplus 1 \rangle - | 1
  \oplus 1 \rangle) = | x \rangle(| 1 \rangle - | 0 \rangle) = -| x
  \rangle(| 0 \rangle - | 1 \rangle)$.}
\begin{equation}
\label{eqn:puredj}
\xrightarrow{U_f} \left (\frac{1}{2^{n/2}}\sum_x^{2^n-1}(-1)^{f(x)}| x
\rangle \right )\left(\frac{| 0 \rangle - | 1 \rangle}{\sqrt
  2}\right).
\end{equation}

If $f$ is constant and $= 0$, this, along with a Hadamard
transformation applied to the first $n$ qubits, will result in:
\begin{align*}
f=0: & & \left (\frac{1}{2^{n/2}}\sum_x^{2^n-1}| x \rangle \right )| -
\rangle \xrightarrow{H^n \otimes I} | 0^n \rangle | - \rangle,
\end{align*}
where $| - \rangle =_{df} \frac{| 0 \rangle - | 1 \rangle}{\sqrt
  2}$. Otherwise if $f$ is constant and $=1$, then this, along with a
Hadamard transformation applied to the first $n$ qubits, will result
in:
\begin{align*}
f=1: & & -\left (\frac{1}{2^{n/2}}\sum_x^{2^n-1}| x \rangle \right )|
- \rangle \xrightarrow{H^n \otimes I} -| 0^n \rangle | - \rangle.
\end{align*}
In either case, a measurement in the computational basis on the first
$n$ qubits yields the bit string $z = 000 \ldots 0 = 0^n = 0$ with
certainty. If $f$ is balanced, on the other hand, then half of the
terms in the superposition of values of $x$ in \eqref{eqn:puredj} will
have positive phase, and half negative. After applying the final
Hadamard transform, the amplitude of $| 0^n \rangle$ will be
zero.\footnote{To illustrate, consider the case where $n=2$. After
  applying $U_f$, the computer will be in the state: $(| 00 \rangle -
  | 01 \rangle + | 10 \rangle - | 11 \rangle)| - \rangle.$ Applying a
  Hadamard transform to the two input qubits will yield:
  \begin{eqnarray*}
  & & \Big((| 00 \rangle + | 01 \rangle + | 10 \rangle + | 11 \rangle) -
  (| 00 \rangle - | 01 \rangle + | 10 \rangle - | 11 \rangle) \\
  & + & (| 00 \rangle + | 01 \rangle - | 10 \rangle - | 11 \rangle) -
  (|00 \rangle - | 01 \rangle - | 10 \rangle + | 11 \rangle)\Big)| -
  \rangle \\
  & = & (0| 00 \rangle + \ldots)| - \rangle.
  \end{eqnarray*}
}
Thus a measurement of these qubits \emph{cannot} produce the bit
string $z = 000 \ldots 0 = 0^n = 0.$ In sum, if the function is
constant, then $z = 0$ with certainty, and if the function is
balanced, $z \neq 0$ with certainty. In either case, the probability
of success of the algorithm is 1, using only a \emph{single}
invocation. This is exponentially faster than any known classical
solution.

\subsection{The necessity of entanglement for explanation thesis}
\label{s:net}

In the literature on quantum computation
(cf. \citealt[]{ekert1998, steane2003}) it is often suggested that
entanglement, such as that present in states like
\eqref{eqn:puredj}, is \emph{required} if a quantum algorithm is
to be capable of achieving a speedup over its classical
alternatives. I will call this the \emph{necessity of an entangled
  state} thesis (NEST). I will call the related claim that
entanglement is a necessary component of any \emph{explanation} for
quantum speedup the \emph{necessity of entanglement for explanation}
thesis (NEXT).\footnote{The attentive reader who has noticed that
  there is actually no entanglement in \eqref{eqn:puredj} when
  $n = 1$ will be somewhat puzzled by both of these theses. In fact,
  as we will see, entanglement will only appear for $n \geq 3$. In
  what follows I will argue, however, that this turns out to be
  evidence for, not against, the NEXT, and indeed does not contradict
  the NEST. This will be clarified in the next section.}

Note that although the NEXT is related to the NEST, these two claims
are not strictly speaking identical. As we will see in \textsection
\ref{s:mixdj}, it is possible for the NEXT to be true even if the
NEST is false (in the technical sense of \textsection
\ref{s:mixent}), and it is not incoherent to argue that the NEXT is
false by citing, as a counter-example, a quantum computer whose state
is always entangled, as we shall see in \textsection
\ref{s:oneq}.

\subsection{De-quantisation}
\label{s:deq}

At first sight the following consideration seems problematic for both
the NEST and the NEXT. Consider the Deutsch-Jozsa algorithm
(cf. \textsection \ref{s:djalgo}) for the special case of
$n=1$. This case is essentially a solution for Deutsch's
problem. Deutsch's \citeyearpar[]{deutsch1985} original solution to
this problem is regarded as the first quantum algorithm ever developed
and as the first example of what has since come to be known as quantum
speedup. If one considers the steps of the algorithm as given in
\textsection \ref{s:djalgo}, however, then the reader can confirm
that, when $n=1$, at no time during the computation are the two qubits
employed actually entangled with one another. The thesis that
entanglement is a necessary condition for quantum speedup thus seems
false. But the situation is not as dark for the NEST and the NEXT as
it appears, since for the case of $n=1$, it is also the case that the
problem can be `de-quantised', i.e., solved just as efficiently using
classical means.

One method for doing this \citep[cf.][]{abbott2010} is with a computer
which utilises the complex numbers $\{1,i\}$ as a computational basis
in lieu of $\{| 0 \rangle, | 1 \rangle\}$. A complex number
$z\in\mathbb{C}$ can be written as $z = a +bi$, where
$a,b\in\mathbb{R}$, and thus can be expressed as a superposition of
the basis elements in much the same way as a qubit.\footnote{Regarding
  the physical realisation of such a computer, note that complex
  numbers can be used, for instance, to describe the impedances of
  electrical circuits and that we can apply the superposition theorem
  to their analysis.} The algorithm proceeds in the following way. We
first note that the action of $U_f$ on the first $n$ qubits in
Eq. \eqref{eqn:puredj} can, for the case of $n=1$, be expressed
as:\footnote{Note that, since $f(0) = f(0)$, $(-1)^{f(0) \oplus f(0)
  \oplus f(1)} = (-1)^{f(1)}$.}
\begin{align*}
& \frac{1}{\sqrt 2}\Big((-1)^{f(0)}| 0 \rangle + (-1)^{f(1)}| 1
\rangle\Big) \nonumber \\
= & \frac{(-1)^{f(0)}}{\sqrt 2}\Big(| 0 \rangle + (-1)^{f(0) \oplus
  f(1)}| 1 \rangle\Big).
\end{align*}
We now define an operator $C_f$, analogously to $U_f$, that acts on a
complex number as follows:
$$C_f(a+bi) = (-1)^{f(0)}\Big(a + (-1)^{f(0) \oplus f(1)}bi\Big).$$
When $f$ is constant, the reader can verify that $C_f(z) = \pm (a+bi)
= \pm z$. When $f$ is balanced, $C_f(z) = \pm(a-bi) = \pm
z^*$. Multiplying by $z/2$ so as to project our output back on to the
computational basis, we find, for the elementary case of $z = 1 + i$,
that
\begin{eqnarray*}
f \mbox{ constant}: & \frac{1}{2}z\cdot\pm z = \pm i\\
f \mbox{ balanced}: & \frac{1}{2}z\cdot\pm z^* = \pm 1.
\end{eqnarray*}
Thus for any $z$, if the result of applying $C_f$ is imaginary, then
$f$ is constant, else if the result is real, then $f$ is balanced
(indeed, by examining the sign we will even be able to tell
\emph{which} of the two balanced or two constant functions $f$
is). This algorithm is just as efficient as its quantum counterpart.

It can similarly be shown \citep[cf.][]{abbott2010} that no
entanglement is present in \eqref{eqn:puredj} when $n=2$, and that
for this case also it is possible to solve the problem efficiently
using classical means.
When $n \geq 3$, however, \eqref{eqn:puredj_uni} is an entangling
evolution and \eqref{eqn:puredj} is an entangled
state. Unsurprisingly, it is no longer possible to define an operator
$C_f$ analogous to $U_f$ that takes product states to product states,
and it is no longer possible to produce an equally efficient classical
counterpart to the Deutsch-Jozsa algorithm \citep[cf.][]{abbott2010}.

Indeed, for the general case, \citeauthor{abbott2010} has shown that a
quantum algorithm can always be efficiently de-quantised whenever the
algorithm does not entangle the input states. Far from calling into
question the role of entanglement in quantum computational speedup,
the fact that the Deutsch-Jozsa algorithm does not require
entanglement to succeed for certain special cases actually provides
(since in these cases it can be de-quantised) evidence for both the
NEST and the NEXT.

\section{Challenges to the necessity of entanglement for explanation
  thesis}
\label{s:mix}

In their own analysis of de-quantisation, \citet{jozsa2003} similarly
find that, for pure quantum states, ``the presence of multi-partite
entanglement, with a number of parties that increases unboundedly with
input size, is necessary if the quantum algorithm is to offer an
exponential speed-up over classical computation.''\footnote{For some
  earlier results relating to specific classes of algorithms, see
  \citet[]{linden2001,braunstein2002}. For a review, see
  \citet[]{pati2009}.} In the same article, however,
\citeauthor[]{jozsa2003} speculate as to whether it may be possible to
achieve exponential speedup, without entanglement, using \emph{mixed}
states. In fact, as we will now see, it is possible to achieve a
modest (i.e., sub-exponential) speedup using unentangled mixed
states. As I will argue, however, entanglement nevertheless plays an
important role in the computational ability of these states, despite
their being unentangled in the technical sense of \textsection
\ref{s:mixent}. Thus, while such counter-examples demonstrate the
falsity of the NEST, they do not demonstrate the falsity of the NEXT.

\subsection{The mixed-state Deutsch-Jozsa algorithm}
\label{s:mixdj}

We will call a `pseudo-pure-state' of $n$ qubits any mixed state that
can be written in the form:
\begin{align*}
\rho_{\mbox{\tiny PPS}}^{\{n\}} \equiv \varepsilon| \psi \rangle\langle
\psi | + (1 - \varepsilon)\mathscr{I},
\end{align*}
where $| \psi \rangle$ is a pure state on $n$ qubits, and
$\mathscr{I}$ is defined as the totally mixed state
$(1/2^n)$I$_{2^n}$. It can be shown that such a state is separable
(cf. \textsection\ref{s:mixent}) and remains so under unitary
evolution just so long as $$\varepsilon < \frac{1}{1 + 2^{2n - 1}}.$$

Now consider the Deutsch-Jozsa algorithm once again (cf. \textsection
\ref{s:djalgo}). This time, however, let us replace the initial
pure state $| 0^n \rangle| 1 \rangle$ with the pseudo-pure state:
\begin{equation}
\label{eqn:djpps}
\rho = \varepsilon| 0^n \rangle| 1 \rangle\langle 0^n |\langle 1 | +
(1 - \varepsilon)\mathscr{I}.
\end{equation}
The algorithm will continue as before, except that this time our
probability of success will not be unity.

To illustrate: assume that the system represented by $\rho$ has been
prepared in the way most naturally suggested by \eqref{eqn:djpps};
i.e., that with probability $\varepsilon$, it is prepared as the pure
state $| 0^n \rangle| 1 \rangle$, and with
probability $1 - \varepsilon$, it is prepared as the completely mixed
state $\mathscr{I}$. Now imagine that we write some of the valid
Boolean functions $f:\{0,1\}^n\to\{0,1\}$ onto balls which we then
place into an urn, and assume that these consist of an equal number of
constant and balanced functions. We select a ball from the urn and
then test the algorithm with this function to see if the algorithm
successfully determines $f$'s type.

Consider the case when $f$ is a constant function. In this case, we
will say the algorithm succeeds whenever it yields the bit string
$z=0$.  We know, from \textsection \ref{s:djalgo}, that the
algorithm will certainly succeed (i.e., with probability 1) when the
system is actually in the pure state $| 0^n \rangle| 1 \rangle$
initially. Given our particular state preparation procedure,
the system is in this state with probability $\varepsilon$. The rest
of the time (i.e., with probability $1 - \varepsilon$), the system is
in the completely mixed state $\mathscr{I}$. In this latter case,
since there are $2^n$ possible values that can be obtained for $z$,
the probability of successfully obtaining $z=0$ will be $1/2^n$. Thus
the overall probability of success associated with the system when $f$
is constant is:
\begin{equation}
\label{eqn:fconstsucc}
P(z=0|f \mbox{ is constant}) = \varepsilon + (1 - \varepsilon)/2^n.
\end{equation}
The probability of failure is:
\begin{equation}
\label{eqn:fconstfail}
P(z \neq 0| f\mbox{ is constant}) = \frac{2^n - 1}{2^n}\cdot(1 -
\varepsilon).
\end{equation}
In the case where $f$ is balanced, a result of $z \neq 0$ represents
success, and the respective probabilities of success and failure are:
\begin{align}
\label{eqn:fbalsucc}
P(z \neq 0|f \mbox{ is balanced}) &= \varepsilon + \frac{2^n
  -1}{2^n}\cdot(1-\varepsilon), \\
\label{eqn:fbalfail}
P(z = 0|f \mbox{ is balanced}) &= (1-\varepsilon)/2^n.
\end{align}

Note that as I mentioned in \textsection\ref{s:mixent}, mixed states
can in general be prepared in a variety of ways. What I have above
called the `most natural' state preparation procedure associated with
\eqref{eqn:djpps}, in particular, is only one of many possible
state preparations that will yield an identical density matrix
$\rho$. For ease of exposition, and in order to see clearly why
Eqs. (\ref{eqn:fconstsucc}-\ref{eqn:fbalfail}) hold, it was
easiest to assume, as I did above, that the system has been prepared
in the way most naturally suggested by \eqref{eqn:djpps}. But
note that there is no loss of generality here; the identities
(\ref{eqn:fconstsucc}-\ref{eqn:fbalfail}) do not depend on the
fact that we have used this particular preparation procedure.

In any case, consider the alternative to the Deutsch-Jozsa algorithm
of performing \emph{classical} function calls on $f$ with the object
of determining $f$'s type. The reader should convince herself that a
single such call, regardless of the result, will not change the
probability of correctly guessing the type of the function $f$. Thus
the amount of information about $f$'s type that is gained from a
single classical function call is zero.\footnote{This information gain
  is referred to as the \emph{mutual information} between two
  variables (in this case, between the type of the function and the
  result of a function call). For more on the mutual information and
  other information-theoretic concepts, see
  \citet[]{nielsenChuang2000}.} On the other hand, as we should expect
given (\ref{eqn:fconstsucc}-\ref{eqn:fbalfail}), for the mixed-state
version of the Deutsch-Jozsa algorithm, it can be shown that the
information gained from a single invocation of the algorithm is
greater than zero for all positive $\varepsilon$, and that this is the
case even when $\varepsilon < \frac{1}{1 + 2^{2n - 1}};$ i.e., the
threshold below which $\rho$ no longer qualifies as an entangled
state. Indeed, this is the case even when $\varepsilon$ is arbitrarily
small \citep[cf.][]{biham2004}, although the information gain in this
case is likewise vanishingly small.

\subsection{Explaining speedup in the mixed-state Deutsch-Jozsa
  algorithm}
\label{s:expmixdj}

The first question that needs to be answered here is whether the
sub-exponential gain in efficiency that is realised by the mixed-state
Deutsch-Jozsa algorithm should qualify as quantum speedup at all. On
the one hand, from the point of view of computational complexity
theory \citep[cf.][]{papadim1994, aaronson2012}, the solution to the Deutsch-Jozsa
problem provided by this algorithm is no more efficient than a
classical solution: from a complexity-theoretic point of view, a
solution $S_1$ to a problem $P$ is deemed to be just as efficient as a
solution $S_2$ so long as $S_1$ requires at most a polynomial increase
in the (time or space) resources required to solve $P$ as compared
with $S_2$. From this point of view, only an \emph{exponential}
reduction in time or space resources can qualify as a true increase in
efficiency. Clearly, the mixed-state Deutsch-Jozsa algorithm does not
yield a speedup over classical solutions, in this sense, when
$\varepsilon$ is small. In fact it can be shown
\citep[1148]{vedral2010} that exponential speedup, and hence a true
increase in efficiency from a complexity-theoretic point of view, is
achievable \emph{only} when $\varepsilon$ is large enough for the
state to qualify as an entangled state.

On the other hand, there is a very real difference, in terms of the
amount of information gained, between one invocation of the black box
\eqref{eqn:djpps} and a single classical function call\textemdash
which is all the more striking since the amount of information one can
gain from a single classical function call is actually zero. Further,
one should not lose sight of the fact that the complexity-theoretic
characterisation of efficient algorithms is artificial and, in a
certain sense, arbitrary. For instance, on the complexity-theoretic
characterisation of computational efficiency, a problem, which for
input size $n$, requires $\approx n^{1000}$ steps to solve is
polynomial in terms of time resources in $n$ and thus tractable, while
a problem that requires $\approx 2^{n/1000}$ steps to solve is
exponential in terms of time resources in $n$ and therefore considered
to be intractable. In this case, however, the `intractable' problem
will typically require much less time to compute than the `tractable'
problem, for all but very large $n$.\footnote{For example, for $n =
  1,000,000$, the easy problem requires $(10^6)^{1000} = 10^{6000}$
  steps to complete while the hard problem requires $2^{1000}$
  steps.} Such extraordinary examples aside, for most practical
purposes the complexity-theoretic characterisation of efficiency is a
good one. Nevertheless it is important to keep in mind that this is a
practical definition of efficiency which does not reflect any deep
mathematical truth or make any deep ontological claim about what is
and is not efficient in the common or pre-theoretic sense of that
term.

But let us come back now from this slight digression to our main
discussion, and let us consider the question of whether entanglement
plays a role in the speedup exhibited by this mixed state. The
strongest argument in favour of a negative answer to this question is,
I believe, the following. Recall that what I have called the `most
natural' state preparation procedure associated with
\eqref{eqn:djpps} is only one of many possible ways to prepare the
system represented by $\rho$. It is possible to prepare the system in
an alternate way if we so desire. Likewise, when $\varepsilon$ is
sufficiently small, it is possible to prepare the final state of the
computer, $\rho_{fin}$, as a mixture of product states. This, in fact, is
the significance of asserting that $\rho_{fin}$ is unentangled. Thus
while the state preparation most naturally suggested by
\eqref{eqn:djpps} may well function as a conceptual tool for
\emph{finding} mixed quantum states that display a computational
advantage (i.e., by enabling a facile derivation of the identities
(\ref{eqn:fconstsucc}-\ref{eqn:fbalfail})), \emph{once found},
it seems as though we may do away with this way of thinking of the
system entirely. Hence there seems to be no need to invoke
entanglement in order to explain the speedup obtainable with this
state.

I believe this line of reasoning to be misleading, however, for it
emphasises the abstract density operator representation of the
computational state at the cost of obscuring the nature of the
computational process that is actually occurring in the computer. To
the point: the density operator corresponding to a quantum system
should not be understood as a representation of the actual physical
state of the system. Rather, the density operator representation of a
quantum system should be understood as a representation of our
knowledge of the space of physical states that the system can possibly
be in, and of our ignorance as to which of these physical states the
system is actually in.

From the point of view of quantum mechanics, it is \emph{pure} states
of quantum systems which should be seen as representations of the
`actual' physical states of such systems, for pure states represent
the most specific description of a system that is possible from within
the theory. I have enclosed the word \emph{actual} within inverted
commas in the preceding sentence in order to emphasise the weakness of
the claim I am making. This claim is not intended to rule out that
there may be a deeper physical theory underlying quantum mechanics,
within which quantum mechanical pure states can be seen as merely
derivative representations. Nor is it intended to rule out that
quantum mechanics only incompletely (as a matter of principle)
specifies the nature of the physical world. I am only making what
should be the uncontroversial claim that relative to quantum mechanics
itself, pure states should be interpreted as those which are most
fundamental, in the sense that they represent a maximally specific
description, within the theory, of the systems in question\textemdash
i.e., they represent the best \emph{grasp} available, from within that
theory, of the real physical situation.

Physics is the science of what is real, in the very minimal sense that
physical concepts \emph{purport} to \emph{give us some idea} of what
the world is like. And if pure states represent the best possible,
i.e., the most specific, representation of the physical situation from
the point of view of a theory, then with right should they be treated
as the more fundamental concepts of the theory. Mixed states, on the
other hand, should be seen as derivative in the sense that they are
abstract characterisations of our knowledge of the space \emph{of pure
  states} a system may be possibly in,\footnote{If one prefers, one
  can think of a mixed state as a \emph{statistical} state,
  representing the mean values of a hypothetical ensemble of
  systems. The difference is inessential to this discussion.} and of
our ignorance of precisely which state within this space the system is
actually in.

If the reader accepts this difference in fundamental
status that I have accorded to pure and mixed quantum
states,\footnote{My claim is intended to be weak enough to be
  compatible with interpretations of the quantum state such as
  Spekkens's, in which quantum states are analogous to the state
  descriptions of his toy theory (cf. \citealt[]{spekkens2007}), in
  that they represent \emph{maximal}, though in principle
  incomplete, knowledge \emph{of} the system in question. It is also
  intended to be compatible with Fuchs's statement that ``... the
  quantum state represents a collection of subjective degrees of
  belief about \emph{something} to do with that system ...''
  \citep[989-990]{fuchs2003}. Nevertheless, the compatibility of my
  claim with Fuchs's and Spekkens's views may be doubted by some. This
  is not the place to attempt to give a reading of either Fuchs's or
  Spekkens's opinions on the interpretation of the quantum state
  description, however. While I may be incorrect as regards the
  compatibility of my claim with their views, I hope that most
  readers will, regardless, appreciate the benign nature of and be
  agreeable to the claim that I am making here. In any case I will be
  assuming it in the remainder of this paper. (For a more
  in-depth treatment of Fuchs's and Spekkens's interpretation of the
  quantum state description, see: \citealt{tait2012}.)} then she
should agree that if it is an explanation of the physical process
actually occurring in the computer that we desire, then it will not do
to limit ourselves to analysing the characteristics of the computer's
`black box' mixed state; rather, we should attempt to give a more
detailed `white box' characterisation of the operation of the computer
in terms of its underlying pure states. Recall the fact\textemdash
which we noted in our earlier discussion of de-quantisation\textemdash
that the unitary evolution \eqref{eqn:puredj_uni} is, in general,
an \emph{entangling   evolution}; i.e., it will take pure product
states, such as, for instance, $| 0^n \rangle| 1 \rangle$, to
entangled states. Now imagine that the computer is initially prepared
in the most natural way suggested by the pseudo-pure state
representation \eqref{eqn:djpps}. Call this `most natural' state
preparation: $s_{ini}$. Imagine further that the computer evolves in
accordance with the entangling unitary transformation $U_f$. This will
yield the transformation
$$
| 0^n \rangle| 1 \rangle \xrightarrow{U_f} | \phi
\rangle
$$
with probability $\varepsilon$, and the transformation
$$
\mathscr{I} \xrightarrow{U_f} \mathscr{I}
$$
with probability $1 - \varepsilon$, where $| \phi \rangle$ is an
entangled state. Thus at the end of the computation, the system will
be in the state $| \phi \rangle$ with probability $\varepsilon$ and in
the state $\mathscr{I}$ with probability $1 - \varepsilon$. Call this
combination of possible states for the system $s_{fin}$. Now at the
end of the computation, the state of the computer will be expressible
by means of the density operator
$$
\rho_{fin} = \varepsilon | \phi \rangle\langle \phi | + (1 -
\varepsilon)\mathscr{I}.
$$
The most natural way that suggests itself for preparing the system
represented by $\rho_{fin}$ is $s_{fin}$. However, one may instead
imagine a state preparation procedure $s'_{fin}$ involving only
product states that would result in an equivalent density operator
representation. Because of this, it is concluded by some that
entanglement plays no role in the computational advantage exhibited by
the computer in this case.

The significance of the fact that $U_f$ is an entangling evolution,
however, is that $s_{ini}$, evolved in accordance with $U_f$, will
\emph{not} result in the combination of states $s'_{fin}$\textemdash
rather, it will result in the combination of states $s_{fin}$. Since
both state preparations, $s_{fin}$ and $s'_{fin}$, yield the same
density matrix representation, they are, from this point of view,
equivalent, but one \emph{cannot} directly obtain $s'_{fin}$ from an
application of $U_f$ to $s_{ini}$.\footnote{I am indebted to Wayne
  Myrvold for suggesting this line of thought, and for helping to
  clear up the conceptual confusions regarding this issue that have
  plagued me to date. I am also indebted to the discussion in
  \citet[\textsection 5]{jozsa2003}. I should note, also, that
  \citet[]{long2002} make a similar point to the one made here; but
  in making it they unnecessarily rely on interpreting the density
  matrix of a system as representing the average values of a
  physical ensemble (i.e. of an actual collection of physical
  systems). The objection is equally forceful, however, whether one
  thinks of the mixed state as representing a physical or a
  statistical ensemble, and whether one thinks of the probabilities
  as ignorance probabilities or as representing relative
  frequencies.}

What of the fact, however, that $\varepsilon$ in the state preparation
$s_{fin}$ may be \emph{vanishingly small} in principle and yet
\emph{still} lead to a computational advantage\textemdash does not
this tell against attributing the speedup exhibited by the computer to
entanglement? I do not believe it does. One must not lose sight of the
fact that ``vanishingly small'' $\neq$ 0. If $\varepsilon$ \emph{were}
actually equal to zero, it is evident that there would, in fact, be no
performance advantage.

It is interesting, nevertheless, to consider the question of what can
happen in the quantum computer when $\varepsilon = 0;$ i.e., when the
state of the computer initially just is the totally mixed state
$\mathscr{I}$. Note that this does not signify that it is impossible
for the computer to actually have been prepared in the pure state $|
0^n \rangle| 1 \rangle$ initially. Rather, it represents the
circumstance where we are \emph{completely ignorant} of the initial
state preparation of the quantum computer; for instance, if the
computer has been prepared as an equally weighted mixture of the basis
states:
\begin{align}
\label{eqn:totallymixed}
\rho_{ini} = \mathscr{I} = \frac{1}{2^n}\sum_{x=0}^{2^n - 1}| x
\rangle\langle x |.
\end{align}
Suppose then, that the quantum computer, represented by the density
operator $\rho_{ini} = \mathscr{I}$, actually is in $| 0^n \rangle| 1
\rangle$ at the start of the computation. Is a computational process
occurring which would enable quantum speedup? From one point of view,
the answer is yes, for the entangling unitary evolution $U_f$ evolves
the computer to an entangled state which is then capable of being
utilised in principle in order to solve the problem under
consideration with fewer computational resources than a classical
computer. In fact, it is not even necessary for the computer to
actually be in the state $| 0^n \rangle| 1 \rangle$ initially to
enable a performance advantage. As long as we know, or at least are
not completely ignorant of, the actual initial pure state of the
computer, any of the basis states can, with suitable manipulation, be
used to obtain a performance advantage.

From another point of view, however, the answer is no, for because we
are completely ignorant as to the actual initial state of the
computer, we will be completely ignorant as to which operation to
perform in order to take advantage of this resource. This sounds
paradoxical, but I think it rather illustrates an important distinction:
between what is actually occurring in a physical system,
on the one hand, and the use which can be made of it by \emph{us}, who
are attempting to achieve some particular end. In the example we are
considering here there assuredly is a process occurring in the
computer that is of the right sort to enable a quantum speedup, but
because we are completely ignorant of the computer's initial
state\textemdash i.e., because there is too much `noise' in the
computer\textemdash we are unable to take advantage of it to achieve
the end of solving the Deutsch-Jozsa problem using fewer computational
resources than a classical computer.

\subsection{DQC1: The power of one qubit}
\label{s:oneq}

In the last subsection we saw that it is possible to achieve a
sub-exponential speedup for the Deutsch-Jozsa problem with an
unentangled mixed-state. We concluded that while this does disprove
the NEST, it does not constitute a counter-example to the NEXT, since
the computational algorithm in question is successful only when the
evolution of the state of the computer is an entangling evolution;
therefore the underlying final state of the computer will always
contain some entanglement despite the fact that the density operator
representation of the final state will be unentangled.

We now consider another purported counter-example to the NEXT. This is
the \emph{deterministic quantum computation with one qubit} (DQC1)
model of quantum computation, which utilises a mixed quantum state to
compute the trace of a given unitary operator and displays an
\emph{exponential} speedup over known classical solutions. As we will
see, the claim sometimes made to the effect that the DQC1 achieves
this speedup without the use of entanglement is unsubstantiated. The
NEXT, however, is not the claim that any state that displays quantum
computational speedup must be entangled. That is the NEST. The NEXT
is, rather, the different claim that entanglement must play a role in
any physical \emph{explanation} of quantum speedup. We saw in the last
section how it is possible for the NEST to be false\footnote{I mean
  false in the technical sense explained in \textsection
  \ref{s:mixent}.} and the NEXT to be true. In this section I will
address the objection that the NEXT is false even if it is the case
that the state of the quantum computer is always entangled. Those
defending such a view claim that another measure of quantum
correlations, \emph{quantum discord}, is far better suited for the
explanatory role. In what follows I will argue that this conclusion is
misguided. Quantum discord is indeed an enormously useful theoretical
quantity for characterising mixed-state quantum computation\textemdash
perhaps even more useful than entanglement. Nevertheless, more than
just pragmatic considerations must be appealed to if one is to make
the case that a particular feature of quantum systems explains quantum
speedup. Thus I will argue that when one looks deeper, and considers
the quantum state from the \emph{multi-partite} point of view, one
finds that entanglement is involved in the production, and even in the
very definition, of quantum discord; indeed, there are some
preliminary indications that quantum discord is, in fact, but a
manifestation of and not conceptually distinct from entanglement.

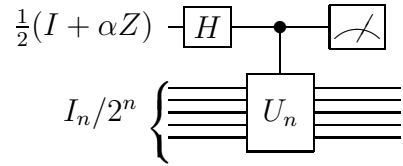
\begin{figure}
  \begin{align*}
  \Qcircuit @C=.5em @R=-.5em {
  & \lstick{\frac{1}{2}(I + \alpha Z)} & \gate{H} & \ctrl{1} & \meter &
  \push{\rule{0em}{4em}} \\
  & & \qw & \multigate{4}{U_n} & \qw & \qw \\
  & & \qw & \ghost{U_n} & \qw & \qw \\
  \lstick{\mbox{$I_n/2^n$}} & & \qw & \ghost{U_n} & \qw & \qw \\
  & & \qw & \ghost{U_n} & \qw & \qw \\
  & & \qw & \ghost{U_n} & \qw & \qw \gategroup{2}{2}{6}{2}{.6em}{\{}
  }
  \end{align*}
  \caption[The DQC1 model]{The DQC1 algorithm for computing the
  trace of a unitary operator.}
  \label{fig:dqc1}
\end{figure}

In the DQC1, or as it is sometimes called: `the power of one qubit',
model of quantum computation \cite[cf.][]{knill1998},\footnote{In this
  exposition of the DQC1, I am closely following \citep[]{datta2005}.} a
collection of $n$ `unpolarised' qubits in the completely mixed state
$I_n/2^n$ is coupled to a single `polarised' control qubit, initialised
to $1/2(I + \alpha Z)$. When the polarisation, $\alpha$, is equal to
1, the control qubit is in the pure state $| 0 \rangle \langle 0 | =
1/2(I + Z)$,
otherwise it is in a mixed state. The problem is to compute the trace
of an arbitrary $n$-qubit unitary operator, $\mbox{Tr}(U_n)$. To
accomplish this, we begin by applying a Hadamard gate to the control
qubit,\footnote{This will yield, for instance, when the control qubit
  is pure, $| 0 \rangle\langle 0 | \xrightarrow{H} \frac{1}{2}\big(|
  0 \rangle\langle 0 | + | 0 \rangle\langle 1 | + | 1 \rangle\langle
  0 | + | 1 \rangle\langle 1 |\big).$} which is then forwarded as part
of the input to a controlled unitary gate that acts on the $n$
unpolarised qubits (see Figure \ref{fig:dqc1}). This results in the
following state for all of the $n+1$ qubits:
\begin{align}
\label{eqn:dqc1}
\rho_{n+1} & = \frac{1}{2^{n+1}}\big(| 0 \rangle\langle 0 | \otimes I_n +
| 1 \rangle\langle 1 | \otimes I_n + \alpha | 0 \rangle\langle 1 |
\otimes U_n^\dagger + \alpha | 1 \rangle \langle 0 | \otimes U_n\big)
\nonumber \\
& = \frac{1}{2^{n+1}}
\left(
\begin{matrix}
  I_n & \alpha U_n^\dagger \\
  \alpha U_n & I_n
\end{matrix}
\right).
\end{align}

The reduced state of the control qubit is
$$\rho_c = \left(
\begin{matrix}
  1 & \alpha\mbox{Tr}(U_n)^\dagger \\
  \alpha\mbox{Tr}(U_n) & 1
\end{matrix}
\right),$$
thus the trace of $U_n$ can be retrieved by applying the $X$ and $Y$
Pauli operators to $\rho_c$. In particular, the expectation values of
the $X$ and $Y$ operators will yield the real and imaginary parts of
the trace, $\langle X \rangle = \mbox{Re}[\mbox{Tr}(U_n)]/2^n$ and
$\langle Y \rangle = -\mbox{Im}[\mbox{Tr}(U_n)]/2^n$, respectively; so
in order to determine, for instance, the real part, we run the circuit
repeatedly, measuring $X$ on the control qubit at the end of each run,
while assuming that the results are part of a distribution whose mean
is the real part of the trace.

Classically, the problem of evaluating the trace of a unitary matrix
is believed to be hard, however for the quantum algorithm it can be
shown that the number of runs required does not scale exponentially
with $n$, yielding an exponential advantage for the DQC1 quantum
computer. When $\alpha < 1$, the expectation values, $\langle X
\rangle$ and $\langle Y \rangle$, are reduced by a factor of $\alpha$
and it becomes correspondingly more difficult to estimate the
trace. However as long as the control qubit has non-zero polarisation,
the model still provides an efficient method for estimating the trace
(and thus an exponential speedup over any known classical solution) in
spite of this additional overhead.

We might ask whether, in a way analogous to the mixed-state
Deutsch-Jozsa algorithm, we can make $\alpha$ small enough so that the
overall state of the DQC1 is demonstrably separable. The answer seems
to be no. On the one hand, for any system of $n+1$ qubits there is a
ball of radius $r$ (measured by the Hilbert-Schmidt norm and centred
at the completely mixed state), within which all states are separable
\citep[]{braunstein1999,gurvits2003}. On the other hand, the state of
the DQC1 is at all times at a fixed distance $\alpha 2^{-(n+1)/2}$
from the completely mixed state. Unfortunately the radius of the
separable ball decreases exponentially faster than $2^{-(n+1)/2}$
\citep[2]{datta2005}.

\begin{figure}
\begin{tabular}{lllllll}
\raisebox{2.5em}{(a)} & \includegraphics[scale=0.25]{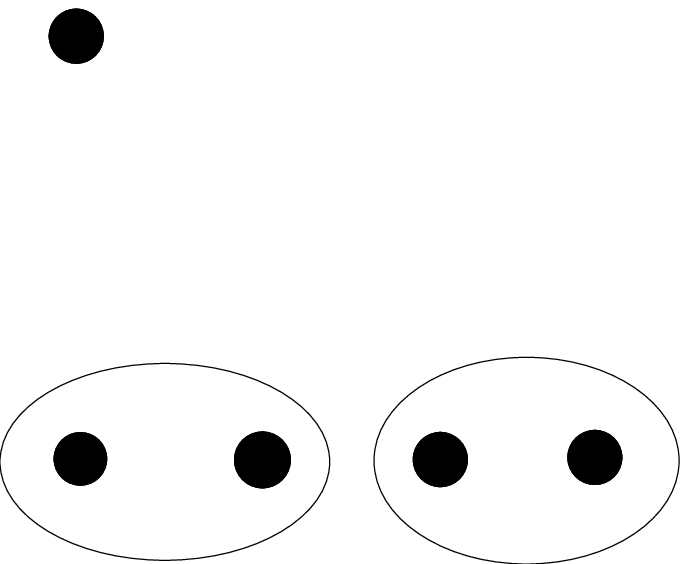} & &
\raisebox{2.5em}{(b)} & \includegraphics[scale=0.25]{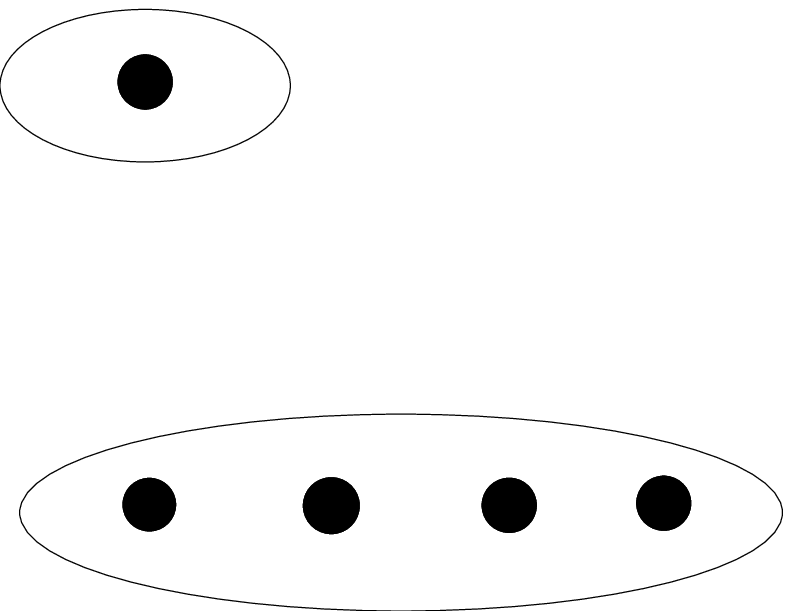}\\
\\
\raisebox{2.5em}{(c)} & \includegraphics[scale=0.25]{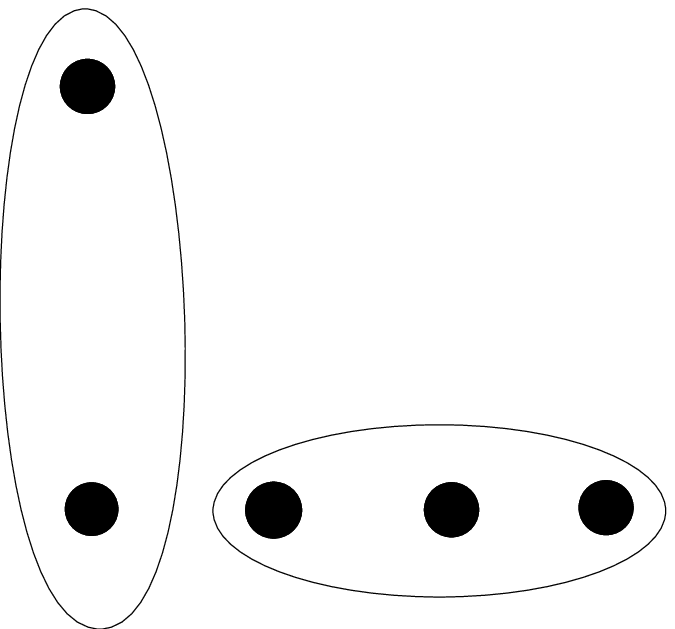} & &
\raisebox{2.5em}{(d)} & \includegraphics[scale=0.25]{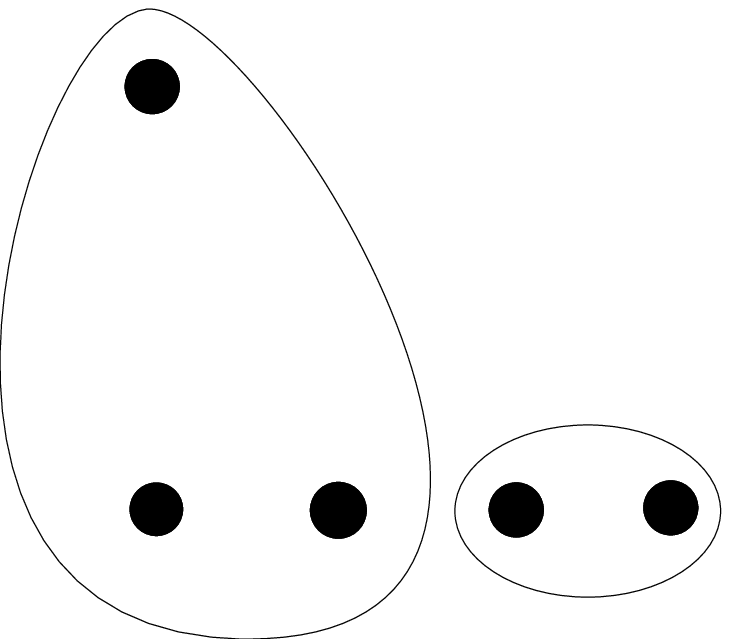} &
\raisebox{2.5em}{(e)} & \includegraphics[scale=0.25]{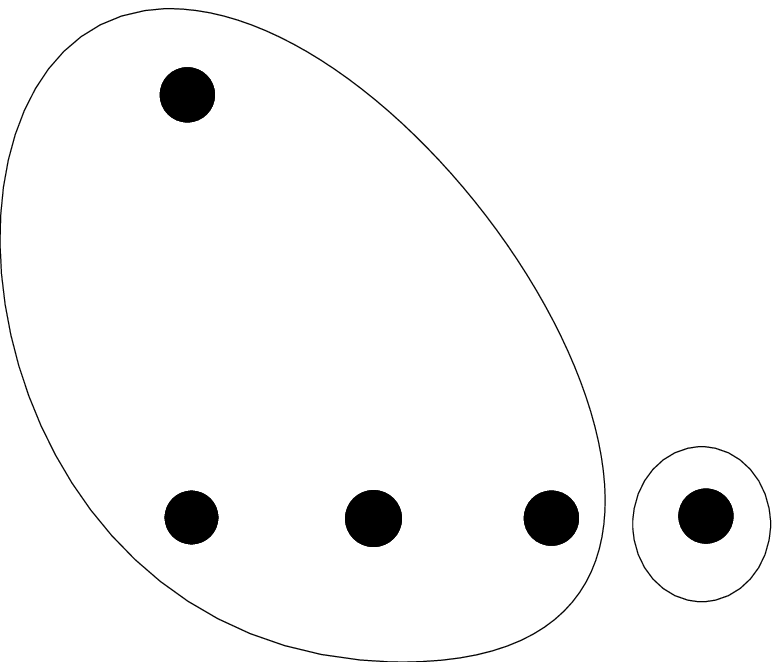}
\end{tabular}
\caption[DQC1 bipartite splits]{Some of the bipartite splits possible
  in the DQC1 for $n=4$. No entanglement can ever occur amongst the
  $n$ unpolarised qubits (a) or between the polarised qubit and the
  rest (b); however, bipartite splits such as (c), (d), and (e) can
  exhibit entanglement \citep[]{datta2005}.}
\label{fig:dqc1_splits}
\end{figure}

Thus, as \citep[2]{datta2005} assert, there appears to be good reason
to suspect that the state \eqref{eqn:dqc1} is an entangled state,
at least for some $U_n$; but it is not obvious where this entanglement
\emph{is}. On the one hand, there is no bipartite entanglement among
the $n$ unpolarised qubits. On the other hand the most natural
bipartite split of the system, with the control qubit playing the role
of the first subsystem and the remaining qubits playing the role of
the second, reveals no entanglement between the two subsystems,
regardless of the choice of $U_n$. When $\alpha > 1/2$, entanglement
can be found when we examine other bipartite divisions amongst the
$n+1$ qubits (see Figure \ref{fig:dqc1_splits}), however, besides
being exceedingly difficult to detect, the amount of entanglement in
the state (as measured by the multiplicative negativity;
cf. \citealt[]{plenio2007}) becomes vanishingly small as
$n$ gets large. Commenting on this circumstance, \citet[13]{datta2005}
write ``This hints that the key to computational speedup might be the
global character of the entanglement, rather than the amount of the
entanglement. ... what happier motto can we find for this state of
affairs than \emph{Multam ex Parvo}, or A Lot out of A Little.''

Others have expressed a different viewpoint on the matter. In fact,
both the DQC1 and the mixed-state version of the Deutsch-Jozsa
algorithm have led many (see for instance, \citealt[]{vedral2010}) to
seriously question whether entanglement plays a necessary role in the
explanation of quantum speedup. The result has been a shift in
investigative focus from entanglement to other types of quantum
correlations. One alternative in particular, \emph{quantum discord}
(which I will explain in more detail shortly), has received much
attention in the literature in recent years
\citep[see, e.g.,][]{merali2011}.

On the one hand, the following facts
all seem to run counter to the NEXT: there is no entanglement in the
DQC1 circuit between the polarised and unpolarised qubits\textemdash
the most natural bipartite split that suggests itself\textemdash
during a computation; tests to detect entanglement along other
bipartite splits in the DQC1 when $\alpha \leq 1/2$ have thus far been
unsuccessful;\footnote{The criterion used by \citet[]{datta2005} to
  detect entanglement is the Peres-Horodecki, or Positive Partial
  Transpose (PPT) criterion \citep[]{peres1996,horodecki1996}. The
  partial transpose of a bipartite system, $\sum_{ijkl}p^{ij}_{kl}|
  i \rangle\langle j |\otimes | k \rangle\langle l |$ acting on
  $\mathcal{H}_A \otimes \mathcal{H}_B$ is defined (with respect to
  the system $B$) as: $$\rho^{T_B} \equiv (I \otimes T)\rho =
  \sum_{ijkl}p^{ij}_{kl}| i \rangle\langle j |\otimes (| k
  \rangle\langle l |)^T = \sum_{ijkl}p^{ij}_{kl}| i \rangle\langle j
  |\otimes | l \rangle\langle k |,$$ where $T$ is the transpose map
  on matrices. The PPT criterion states that, if $\rho$ is a
  separable state, then the partial transpose of $\rho$ has
  non-negative eigenvalues. Satisfying the PPT criterion is a
  necessary (but not sufficient) condition for the joint density
  matrix of two systems to be separable. While
  \citeauthor[]{datta2005} were unable to detect entanglement in the
  DQC1 (along any bipartite split) for the case of $\alpha \leq
  1/2$, they nevertheless note that it is very likely that both
  entanglement and bound entanglement are present in the state. A
  state exhibits \emph{bound entanglement} \citep[cf.][]{hyllus2004}
  when, in spite of the fact that it is entangled, no pure entangled
  state can be obtained from it by means of LOCC operations. One
  important characteristic of bound entangled states is that they
  (at least sometimes) satisfy the PPT criterion despite the fact
  that they are entangled.
} and finally, even when $\alpha$ is relatively large, only a
vanishingly small amount of entanglement can be found in the state of
the DQC1 \eqref{eqn:dqc1}. On the other hand, when we consider the
correlations between the polarised and unpolarised qubits from the
point of view of \emph{quantum discord}, it turns out that the discord
at the end of the computation is \emph{always} non-zero along this
bipartite split for \emph{any} $\alpha > 0$
\citep[]{datta2008}. \citet[4]{datta2008} therefore write, \emph{and I
  agree}, that ``for some purposes, quantum discord might be a better
figure of merit for characterizing the quantum resources available to
a quantum information processor.'' All the same, as I will argue
below, it is a mistake to conclude as they and others do that the NEXT
is false; i.e., that entanglement may play no role in the explanation
of the quantum speedup of the DQC1
\citep[]{datta2008,vedral2010,merali2011}; for the NEXT \emph{is
  compatible} with all of these facts.

\subsection{Quantum discord}

Quantum discord \citep[]{henderson2001, ollivier2002}\footnote{Quantum
  discord was introduced independently by both
  \citeauthor[]{henderson2001} and by \citeauthor[]{ollivier2002},
  with slight differences in their respective formulations
  (\citeauthor[]{henderson2001} consider not just projective
  measurements but positive operator valued measures more
  generally). These and other alternative formulations of quantum
  discord do not differ in essentials. The definition of discord I
  introduce here is \citeauthor[]{ollivier2002}'s.} quantifies the
difference between the quantum generalisations of two classically
equivalent measures of mutual information,\footnote{See
  \citet[]{nielsenChuang2000} for an overview of the basic concepts of
  classical and quantum information theory.}
\begin{align}
\label{eqn:clasmut1}
\mathcal{I}_c(A:B) & = H(A) + H(B) - H(A,B), \\
\label{eqn:clasmut2}
\mathcal{J}_c(A:B) & = H(A) - H(A|B).
\end{align}
These two expressions are not equivalent quantum mechanically, for
while \eqref{eqn:clasmut1} has a straightforward quantum
generalisation in terms of the von Neumann entropy $S$:
\begin{align}
\label{eqn:quantmut1}
\mathcal{I}_q(A:B) & = S(A) + S(B) - S(A,B),
\end{align}
things are more complicated for the quantum generalisation of
\eqref{eqn:clasmut2}. The quantum counterpart, $S(A|B)$, to the
conditional entropy requires a specification of the information
content of $A$ given a determination of the state of $B$. Determining
the state of $B$ requires a measurement, however, which requires the
choice of an observable. But in quantum mechanics observables are, in
general, non-commuting. Thus the conditional entropy will be different
depending on the observable we choose to measure on $B$. If, for
simplicity, we consider only perfect measurements, represented by a
set of one dimensional projection operators, $\{\Pi_j^B\}$, this
yields, for the quantum version of \eqref{eqn:clasmut2}, the
expression:
\begin{align}
\label{eqn:quantmut2}
\mathcal{J}_q(A:B) & = S(A) - S(A|\{\Pi_j^B\}).
\end{align}
We now define discord as the minimum value (taken over $\{\Pi_j^B\}$)
of the difference between \eqref{eqn:quantmut1} and
\eqref{eqn:quantmut2}:
\begin{align}
\label{eqn:discord}
\mathcal{D}(A,B) \equiv \mbox{min}_{\{\Pi_j^B\}}\mathcal{I}_q(A:B) -
\mathcal{J}_q(A:B).
\end{align}
Discord is, in general, non-zero for mixed states, while for pure
states it effectively becomes a measure of entanglement
\citep[3]{datta2008}; i.e., for pure states it is equivalent to the
entropy of entanglement \citep[cf.][]{plenio2007}.

Interestingly, there are some mixed states which, though
\emph{separable}, exhibit non-zero quantum discord. For instance,
consider the following bipartite state:
\begin{align}
\label{eqn:discexamp}
\rho_{\mbox{\tiny disc}} = \frac{1}{2}(| 0 \rangle\langle 0 |_A
\otimes | 0 \rangle\langle 0 |_B) + \frac{1}{2}(| 1 \rangle\langle 1
|_A \otimes | + \rangle\langle + |_B).
\end{align}
This state is obviously separable. Since $| 0 \rangle$ and $| +
\rangle$ are non-orthogonal states, however, $\mathcal{J}_q(A:B)$ will
yield a different value depending on the experiment performed on
system $B$; and thus this state will yield a non-zero quantum
discord. Note that this is impossible for a classical state:
classically, it is \emph{always} possible to prepare a state as a
mixture of \emph{orthogonal} product states.

In most of the literature on this topic, one is introduced to quantum
discord as a quantifier of the non-classical \emph{correlations}
present in a state which are not necessarily identifiable with
entanglement. Such an interpretation of the significance of this
quantity is supported by the fact that, in the classical scenario at
least, the mutual information contained in a system of two random
variables is held to be representative of the extent of the
correlations between them. Since the quantum generalisations of the
two classically equivalent measures of mutual information
$\mathcal{I}_c(A:B)$ and $\mathcal{J}_c(A:B)$ are not equivalent,
then, this is taken to represent the presence of non-classical
correlations over and above the classical ones, some, but not all of
which may be accounted for by entanglement, and some by `quantum
discord'.

Interpreting discord as a type of non-classical correlation is
nevertheless puzzling. Consider, for instance, a classically
correlated state represented by the following probability
distribution:
\begin{align}
\label{eqn:strcl}
\frac{1}{2}([+]_l,[+]_r) + \frac{1}{2}[-]_l[-]_r.
\end{align}
Here, let $[\cdot]_l$ represent the circumstance that Linda (in
Liverpool) finds a letter in her mailbox today containing a piece of
paper on which is inscribed the specified symbol ($+$ or $-$), and let
$[\cdot]_r$ represent the occurrence of a similar circumstance for
Robert (in Ravenna). According to the probability distribution, it is
equally likely that they both receive a letter today inscribed with
$+$ as it is that they both receive one inscribed with $-$, but it
cannot happen that they each today receive letters with non-matching
symbols. These correlations are easily explainable classically, of
course. It so happens that yesterday I flipped a fair coin. I observed
the result of the toss and accordingly jotted down either $+$ or $-$
on a piece of paper, photocopied it, and sent one copy each to Robert
in Ravenna and Linda in Liverpool (by overnight courier, of course).

A quantum analogue for classically correlated states such as
\eqref{eqn:strcl} is a mixed state decomposable into product
states:
\begin{align}
\label{eqn:qucl}
\sum_{ij} p_{ij} | i \rangle\langle i | \otimes | j \rangle\langle j |
\end{align}
such that the $| i \rangle$ and $| j \rangle$ are mutually orthogonal
sub-states of the first and second subsystem, respectively. For such a
state it is easy to provide a `hidden variables' explanation, similar
to the one above, that will account for the observed probabilities of
joint experiments on the two subsystems.

We can equally give such a local hidden variables account of the
discordant state $\rho_{\mbox{\tiny disc}}$: tossing a fair coin, I
prepare the state $| 0 \rangle\langle 0 |_A \otimes | 0 \rangle\langle
0 |_B$ if the coin lands heads, and $| 1 \rangle\langle 1 |_A \otimes
| + \rangle\langle + |_B$ if it lands tails. Let $Pr(X, Y | a, b,
\lambda)$ refer to the probability that Alice's $a$-experiment and
Bob's $b$-experiment determine their qubits to be in states $X$ and
$Y$, respectively, given that the result of the coin toss is
$\lambda$. Then (omitting bras and kets for readability):
\begin{align*}
Pr(0, 0 | \hat{z}, \hat{z}, H) & = Pr(0, \cdot | \hat{z}, \cdot, H)
\times Pr(\cdot, 0 | \cdot, \hat{z}, H) = 1, \\
Pr(1, 1 | \hat{z}, \hat{z}, T) & = Pr(1, \cdot | \hat{z}, \cdot, T)
\times Pr(\cdot, 1 | \cdot, \hat{z}, T) = 1/2, \\
Pr(0, + | \hat{z}, \hat{x}, H) & = Pr(0, \cdot | \hat{z}, \cdot, H)
\times Pr(\cdot, + | \cdot, \hat{x}, H) = 1/2, \\
Pr(1, + | \hat{z}, \hat{x}, T) & = Pr(1, \cdot | \hat{z}, \cdot, T)
\times Pr(\cdot, + | \cdot, \hat{x}, T) = 1,
\end{align*}
and so on. More generally, $Pr(X,Y | a,b, \lambda) = Pr(X, \cdot | a,
\cdot, \lambda) \times Pr(\cdot, Y | \cdot, b, \lambda)$. Thus once we
specify the value of $\lambda$ there are no remaining correlations in
the system and the probabilities for joint experiments are
\emph{factorisable}. This should be unsurprising. Given a
specification of $\lambda$, the state of the system is in a
\emph{product state}, after all.

Contrast this with an entangled quantum system such as, for instance,
the one represented by the pure state $$|\Phi^+\rangle = \frac{| 00
  \rangle + | 11 \rangle}{\sqrt{2}}.$$ Bell's theorem
\citep[]{bell1964} demonstrates that the correlations between
subsystems present in such a state cannot be reproduced by any local
hidden variables theory in the manner described above. These
correlations are non-classical.

There is certainly \emph{something} non-classical about a state such
as $\rho_{\mbox{\tiny disc}}$; viz., a quantum state such as
$\rho_{\mbox{\tiny disc}}$, though separable, cannot be prepared as a
mixture of orthogonal product states. Yet it is always possible to so
prepare classical states. As a result, the information one can gain
about Alice's system through an experiment in the $\{+, -\}$ basis on
Bob's system will be different from the information one can gain about
Alice's system through an experiment in the computational basis on
Bob's system. On the one hand, in the absence of a specification of a
hidden parameter such as $\lambda$, given an experiment on $B$ in the
computational basis which determines $B$ to be in state $| 0 \rangle$,
it is still unclear, because of the way in which system $B$ was
prepared, whether the joint system is in the state $| 0 \rangle
\otimes | 0 \rangle$ or in the state $| 1 \rangle \otimes | +
\rangle$. Given an experiment on $B$ in the $\{+, -\}$ basis which
yields $| + \rangle$, on the other hand, it is perfectly clear which
product state the joint system is in. But these facts by themselves
are certainly not indicative of the presence of non-classical
\emph{correlations} between the two subsystems.

There is one indirect sense, however, in which $\rho_{\mbox{\tiny
  disc}}$ can be said to contain non-classical correlations. Recall
from \textsection \ref{s:purify} that any mixture can be
considered as the result of taking the partial trace of a pure
entangled state on a larger Hilbert space. Given that, as I argued in
\textsection \ref{s:expmixdj}, the pure state representation of a
quantum system should be taken as fundamental, we can consider the
bipartite state $\rho_{\mbox{\tiny disc}}$ as in reality but a partial
representation of a tripartite entangled quantum system, where the
third party is an external environment with enough degrees of freedom
to purify the overall system. And since entangled systems do not admit
of a description in terms of local hidden variables, it follows that
the system partially represented by $\rho_{\mbox{\tiny disc}}$ can
legitimately be said to contain non-classical correlations.

Even so it is unclear how these non-classical correlations \emph{per
  se} can have anything to do with the quantum discord exhibited by
$\rho_{\mbox{\tiny disc}}$, for it is also the case that a classically
correlated mixture of \emph{orthogonal} product states, i.e. one of
the form \eqref{eqn:qucl}, can be purified in just the same way as
a discordant one and hence also the case that it can be given a
multi-partite representation in which entanglement is present.

As we will now see, however, there is in fact a tight relationship
between the \emph{amount} of discord associated with a bipartite mixed
state and the \emph{amount} of entanglement associated with a
tripartite representation of that state. And, interestingly from our
point of view, what emerges from this is a correspondingly tight
relationship between the quantum speedup exhibited by the DQC1 and the
amount of entanglement associated with its purified tripartite
representation, and thus a confirmation, not a refutation, of the
NEXT.

\subsection{Explaining speedup in the DQC1}

Quantum discord was introduced independently by
\citeauthor[]{henderson2001} and by \citeauthor[]{ollivier2002} in
\citeyear[]{henderson2001} and \citeyear[]{ollivier2002},
respectively; however, it was only recently given an operational
interpretation, independently by \citet[]{madhok2011} and by
\citet[]{cavalcanti2011}.\footnote{I present here the definition
  given by \citeauthor[]{cavalcanti2011}, although the conclusion I
  will draw is the same regardless of which definition is used.} On
both characterisations, quantum discord is operationally defined in
terms of the entanglement consumed in an extended version of the
quantum state merging protocol \citep[cf.][]{horodecki2005}.

In the quantum state merging protocol, three parties: Alice, Bob, and
Cassandra, share a state $| \psi_{ABC} \rangle$. Quantum state merging
characterises the process, $$| \psi_{ABC} \rangle \to | \psi_{B'BC}
\rangle,$$ by which Alice effectively transfers her part of the system
to Bob while maintaining its coherence with Cassandra's part. It turns
out that in order to effect this protocol a certain amount of
entanglement must be consumed (quantified on the basis of the quantum
conditional entropy, $S(A|B)$; cf. \citealt[]{nielsenChuang2000}.). When we
add to this the amount of entanglement needed (as quantified by the
entanglement of formation; cf. \citealt[]{plenio2007}) to
prepare the state $| \psi_{ABC} \rangle$ to begin with, the result is
a quantity identical to the quantum discord between the subsystems
belonging to Alice and Cassandra at the time the state is prepared.

\begin{figure}
\begin{tabular}{lllllll}
\raisebox{2.5em}{(a)} & \includegraphics[scale=0.3]{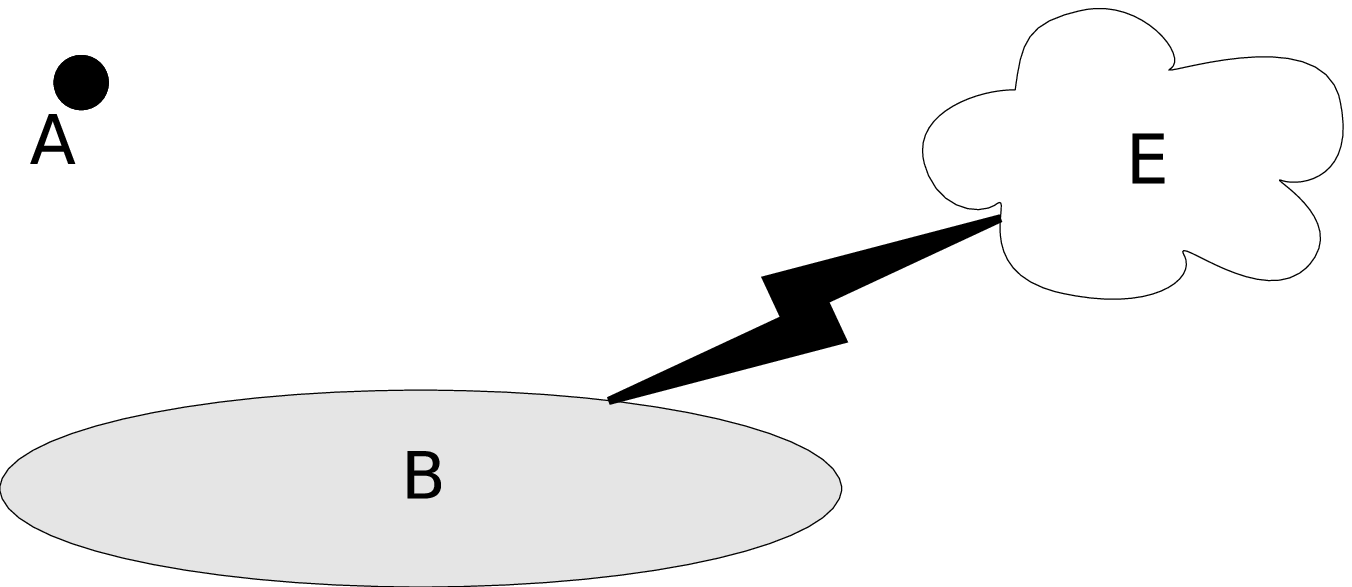} & &
\raisebox{2.5em}{(b)} &
\raisebox{-0.5em}{\includegraphics[scale=0.3]{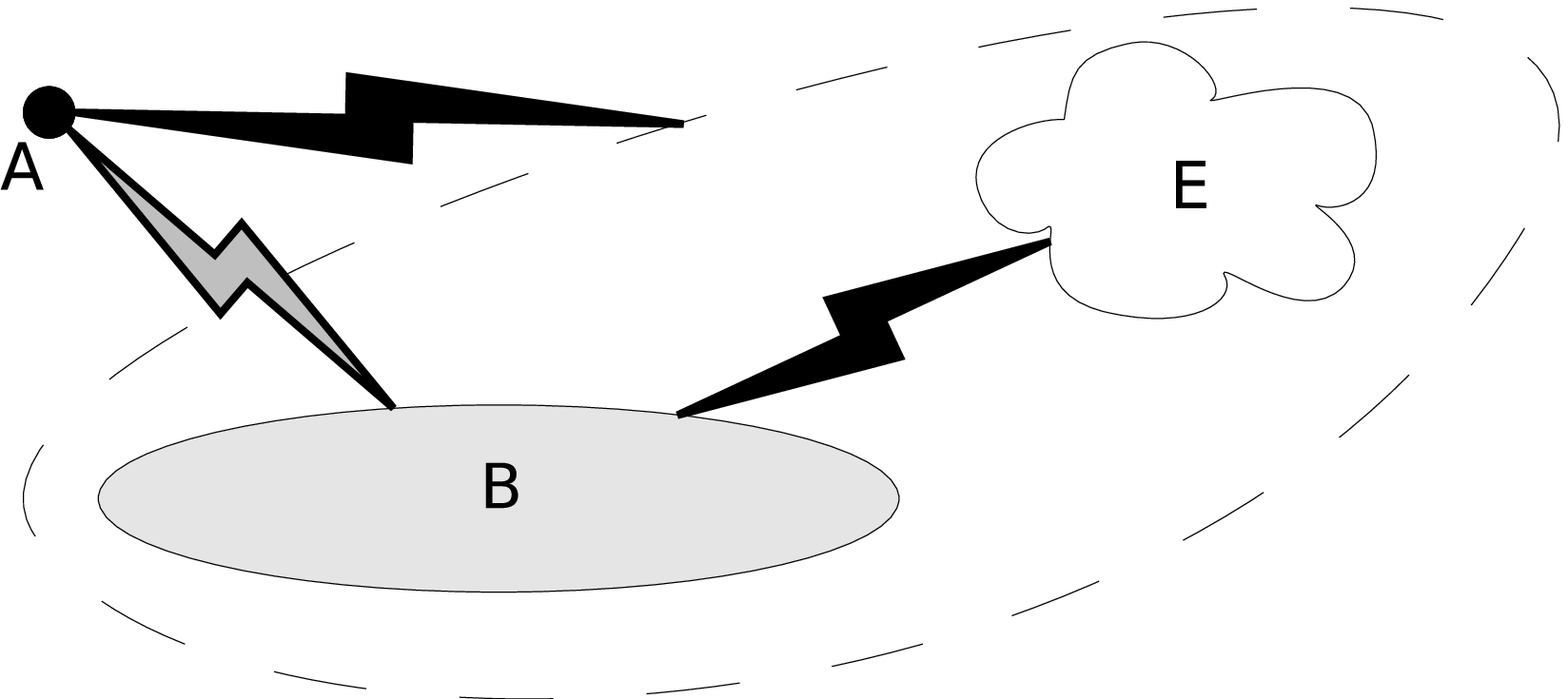}}\\
\end{tabular}
\caption[Tripartite representation of the DQC1]{A (pure) tripartite
  representation of the elements of the DQC1 protocol before (a) and
  after (b) the application of the controlled not gate. Black and
  grey thunderbolts represent entanglement and discord,
  respectively. After the application of the controlled not gate,
  there is an increase in the discord between $A$ and $B$ and a
  corresponding increase in the entanglement between $A$ and the
  combined system $BE$.}
\label{fig:tridqc1}
\end{figure}

The foregoing operational interpretation of discord has an affinity
with an illuminating analysis of the DQC1 circuit due to
\citet[]{fanchini2011}. \citeauthor[]{fanchini2011} show that a
relationship between quantum discord and entanglement emerges when we
consider the DQC1 circuit, not as a bipartite system composed of
polarised and unpolarised qubits respectively, but as a tripartite
system in which the environment plays the role of the third
subsystem. \citeauthor[]{fanchini2011} note that an alternate way of
characterising the completely mixed state of the unpolarised qubits,
$I_n/2^n$, is to view it as part of a bipartite entangled state, with
the second party an external environment having enough degrees of
freedom to purify the overall system. This yields a tripartite
representation for the DQC1 circuit as a whole (see Figure
\ref{fig:tridqc1}).

\citeauthor[]{fanchini2011} show that, for an arbitrary tripartite
pure state, there is a conservation relation between entanglement of
formation and quantum discord. In particular, the sum of the bipartite
entanglement that is shared between a particular subsystem and the
other subsystems of the system cannot be increased without increasing
the sum of the quantum discord between this subsystem and the other
subsystems as well (and vice versa). In the DQC1, after the
application of the controlled not gate (see Figure \ref{fig:dqc1}),
there is an increase in the quantum discord between $B$ and $A$. This
therefore necessarily involves a corresponding increase in the
entanglement between $A$ and the combined system $BE$. All of this
accords with what we would expect given the above operational
interpretation of quantum discord: an increase in quantum discord
requires an increase in the entanglement available for consumption in
a potential quantum state merging process.

Note also that from this tripartite point of view, there is just as
much entanglement in the circuit as there is discord; in particular,
exactly as for quantum discord, there is entanglement in the circuit
whenever it displays a quantum speedup, i.e., for any $\alpha > 0$.

\citeauthor[]{fanchini2011} speculate that it is not the presence of
entanglement or discord (however the latter is interpreted) per se
that is necessary for the quantum speedup of the DQC1, but rather the
ability of the circuit to \emph{redistribute} entanglement and
discord. This thought seems to be confirmed by a theoretical result of
\citet[]{brodutch2011}, who show that shared entanglement is required
in order for two parties to bilocally implement\footnote{Bilocal
  implementation means, in this context, an implementation in which
  Alice and Bob are limited to local operations and classical
  communications \citep[cf.][]{plenio2007}.} \emph{any} bipartite
quantum gate\textemdash even one that operates on a restricted set
$\mathcal{L}$ of unentangled input states and transforms them into
unentangled output states. This means, in particular, that
entanglement is required in order to implement a gate that changes the
discord of a quantum state.

By themselves, these considerations already amount to confirmations of
the NEXT, for entanglement appears to be involved in the very
definition of discord, and it appears that we require entanglement
even for the production of discord in a quantum circuit. But in
addition, there are indications that quantum discord need not be
appealed to at all to give an account of quantum speedup (though such
a characterisation will of course be less practical, as I have already
mentioned), in light of one other recent theoretical
result. \citeauthor{devi2011} \citeyearpar[]{devi2008,devi2011} have
pointed out that more general measurement schemes than the positive
operator valued measures (POVM) used thus far exist for characterising
the correlations present in bipartite quantum systems.

POVMs are associated with completely positive maps and are well suited
for describing the evolution of a system when we can view the system
as uncorrelated with its external environment. When the system is
initially correlated with the environment, however, the reduced
dynamics of the system may, according to \citeauthor{devi2011}, be
`not completely positive'. But as \citeauthor{devi2011} show, from the
point of view of a measurement scheme that incorporates not completely
positive maps in addition to completely positive maps, all quantum
correlations reduce to entanglement.

In sum, it is, I believe, unsurprising that on the standard analysis
the DQC1 circuit displays strange and anomalous correlations in the
form of quantum discord, for the DQC1 is typically characterised as a
\emph{bipartite} system, and from the point of view of a measurement
framework that incorporates only completely positive maps. As
\citeauthor[]{fanchini2011} have shown, however, the DQC1 circuit is
more properly characterised, not as an isolated system, but as a
system initially correlated with an external environment. The
evolution of such a system is best captured by a measurement framework
incorporating not completely positive maps, and within such a
framework, the anomalous correlations disappear and are subsumed under
entanglement. From this point of view the equivalence of entanglement
and discord for pure bipartite states is also unsurprising, for it is
precisely pure states for which the correlation with the environment
can be ignored and for which a framework incorporating only completely
positive maps is appropriate.

The use of not completely positive maps to characterise the evolution
of open quantum systems is not wholly without its detractors. The
question of whether such not completely positive maps are `unphysical'
is an interesting and important one, though I will not address it
here.\footnote{For a more detailed discussion, and qualified defence
  of the use of not completely positive maps, see
  \citet[]{cuffaro2012a}.} But regardless of the answer to this
question, it should be clear, even without the appeal to this more
general framework, that entanglement has \emph{not} been shown to be
unnecessary for quantum computational speedup. Far from being a
counter-example to the NEXT, the DQC1 model of quantum computation
rather serves to illuminate the crucial role that entanglement plays
in the quantum speedup displayed by this computer.

\section{Conclusion}

Quantum entanglement is considered by many to be a necessary resource
that is used to advantage by a quantum computer in order to achieve a
speedup over classical computation. Given \citeauthor[]{jozsa2003}'s
and \citeauthor[]{abbott2010}'s general results for pure states, and
given that, as I argued in \textsection \ref{s:expmixdj}, a pure
state should be considered as the most fundamental representation of a
quantum system possible in quantum mechanics, the burden is upon those
who deny the NEXT to either produce a counter-example or to show, in
some other more principled way, why the view is false. We examined two
such counter-examples in this paper. Upon closer examination we
found neither of these, neither the sub-exponential speedup of the
unentangled mixed-state version of the Deutsch-Jozsa algorithm, nor
the exponential speedup of the DQC1 model of quantum computation,
demonstrate that entanglement is unnecessary for quantum speedup; they
rather make clearer than before the role that entanglement \emph{does}
play, and point the way to a fuller understanding of both entanglement
and quantum computation.

\bibliographystyle{apa-good}
\bibliography{Bibliography}{}

\end{document}